\documentclass[mathleft
]{an}

\usepackage{graphicx}
\usepackage{times}
\newcommand{\citepp}[1]{
  (\cite{#1})
}

\begin{document}

\Pagespan{789}{}
\Yearpublication{2006}%
\Yearsubmission{2006}%
\Month{10}%
\Volume{999}%
\Issue{88}%

\title{Time Distance Study of Isolated Sunspots}

\author{Sergei Zharkov
\thanks{Corresponding author:
  \email{s.zharkov@sheffield.ac.uk}\newline}
\and  Christopher J. Nicholas 
\and Michael J. Thompson 
}
\titlerunning{Time Distance Study of Isolated Sunspots} 
\authorrunning{S. Zharkov {\em et al}}
\institute{
Solar Physics and upper-Atmosphere Research Group, 
          Department of Applied Mathematics, \\
          University of Sheffield, 
          Hounsfield Road, Hicks  
          Building, Sheffield, S3 7RH, UK \\
}

\received{30 May 2005}
\accepted{
11 Nov 2005
}
\publonline{later}


\abstract{%
We present a comparative seismic study of conditions around and beneath isolated sunspots. Using the European Grid of Solar Observations' Solar Feature Catalogue of sunspots derived from SOHO/MDI continuum and magnetogram data, 1996-2005, we identify a set of isolated sunspots by checking that within a Carrington Rotation there were no other spots detected in the vicinity. We then use level-2 tracked MDI Dopplergrams available from SOHO website 
to investigate wave-speed perturbations of such sunspots using time-distance helioseismology.
}

\maketitle
\hyphenation{Fea-tu-re Ca-ta-lo-gues}
\hyphenation{tra-vel ti-me}
\hyphenation{cross-cor-re-lati-on}
\section{Introduction}
Local helioseismology provides a powerful diagnostic of the solar interior, allowing us to infer from observations 
of acoustic waves the structures and flows beneath the Sun's surface. 
In this work we focus on investigation of wave-speed anomalies underneath several $\alpha$ and $\beta$ 
sunspots selected using the European Grid of 
Solar Observations Feature Catalogues \citepp{vzhark05}. 

Previous estimates of sunspot properties from travel-time measurements indicate a negative perturbation 
to the sound speed directly underneath sunspots up to the depths $2-5$ Mm, \citepp{Kosovichev2000}. 
Beneath this depth the sound speed perturbation is positive.

\begin{table*}
\begin{center}
\begin{tabular}{|c|c|c|c|c|c|}
\hline
NOAA \# & AR Type & Date \& Time of Observation & Longitude & Latitude & Area \\
\hline
7973 & $\alpha$ & 1996-06-24T13:05:35.559 & 43.526 & 9.196 & 2.37\\
8038 & $\alpha$ & 1997-05-11T07:08:35.354 &  140.854 & 20.633 & 3.01 \\
9056 & $\alpha$ & 2000-06-26T06:23:32.626 & 177.125 &  -13.113 & 7.140 \\
9138 & $\alpha$ & 2000-08-24T05:59:31.318 & 117.515 & -30.731 & 1.44 \\
9142 & $\alpha$ & 2000-08-27T17:35:31.359 & 46.371 & 15.975 & 1.41\\
9779 & $\beta$ & 2002-01-13T06:23:30.128 & 269.551 & 28.526 & 4.31 \\
\hline
\end{tabular}
\caption{Sunspots used in this study listed by NOAA numbers. Date, Carrington longitude, latitude and area
are taken from the EGSO SFC; times are selected to be closest to the available Dopplergram data. Area is 
measured in heliographic degrees squared. Carrington longitude and latitude are given in degrees.}
\end{center}
\end{table*}

\section{Data and method}

We have used MDI level-2 tracked Dopplergram data. Using the EGSO Sunspot Solar Feature catalogue we have identified a number of sunspots that have no other sunspots within a radius of 10 heliographic degrees. Locating the available MDI level-2 data we have selected 6 sunspots comprising Active Regions NOAA 7973, NOAA 8038, NOAA 9056, NOAA 9138, NOAA 9142, NOAA 9779. From MDI data we have extracted 512-minute long Dopplergram datacubes that were processed using time-distance helioseismology methods. In addition, to investigate the solar activity in these active regions we use EGSO SFC sunspot data 
in conjunction with level-1.8 MDI Continuum and Magnetogram daily synoptic images and NOAA data available via Solar Monitor ({\em  http://www.solarmonitor.org}). Some basic statistics available for the regions are presented in Table 1, while information on evolution of each region can be found in Table 2 (see Section 3 for more details).

\subsection{Time-distance method}
The raw data are measured in the form of Dopplergrams representing line-of-sight velocity in the lower solar atmosphere. The pre-processing stage consists of applying high-pass and f-mode filters \citepp{GB2005} to remove supergranulation and f-mode data. Next in order to select specific acoustic wave-packets and obtain a clear cross-correlation signal for p-modes phase-speed filtering is applied corresponding to a number of  skip-distances. 
Cross-correlation of solar Dopplergram data at two points over time \citepp{Duval93} produces wave-packet like structures, corresponding to the propagation between the two points of packets formed by acoustic waves inside the Sun. 

To avoid stochastic noise, averaging is necessary for obtaining good cross-correlations. In this work, we proceed by computing the cross-correlation of a central point with a surrounding annulus. The annulus thickness is taken to be around $4.5$ Mm. 
We average individual point-to-point cross-correlations $C({\bf r_1}, {\bf r_2}, t)$ over an annulus centered on $\bf{r_1}$ and with a radius $\Delta=|{\bf r_1}-{\bf r_2}|$. Thus we estimate a point-to-annulus cross-correlation function $C({\bf r}, \Delta, t)$ \citepp{Cv05}. 
\subsubsection{Travel Times}
From point-to-annulus cross-correlation 
we derive the one-way travel times $\tau_{i/o}({\bf r}, \Delta)$ of wavepackets propagating inward, $i$, from the annulus to its center and outward, $o$, by fitting separately the negative and positive time-lag portions of the cross-correlation with 
Gabor wavelets \citepp{Kosovichev}. We define the mean travel-times $\tau({\bf r}, \Delta)$ as 
 $\tau({\bf r}, \Delta)=(\tau_i({\bf r}, \Delta)+\tau_{o}({\bf r}, \Delta))/2$.
 
 As a first approximation, perturbations $\delta\tau({\bf r}, \Delta)$ in these mean travel-times are linearly related to the sound-speed perturbations $\delta c$ in the wave propagation region, 
\begin{eqnarray}
\delta \tau( \vec{r}, \Delta) = \int \int_S d{\bf r'} 
\int_{-d}^0 dz K({\bf r- r'}, z;\Delta)\frac{\delta c^2}{c^2}({\bf r'}, z),
\label{eq_TT_Kernel}
\end{eqnarray}
where $S$ is the area of the region, $d$ is its depth and $\delta\tau({\bf r}, \Delta)$ is defined as the difference between the measured travel-time at a given location ${\bf r}$ on the solar surface and the average of the travel-times in the quiet sun. The sensitivity kernel for the relative squared sound speed perturbation is given by $K$. 

In this work we use wave-speed sensitivity kernels estimated using the Rytov approximation \citepp{jensen}. The Rytov and ray approximations give very similar results, with former being more reliable inverting deeper structures 
\citepp{Cv04}.
\subsubsection{Inversion}

\begin{figure*}
\begin{center}
\includegraphics[height=12.5cm, width=8.cm, angle=90]{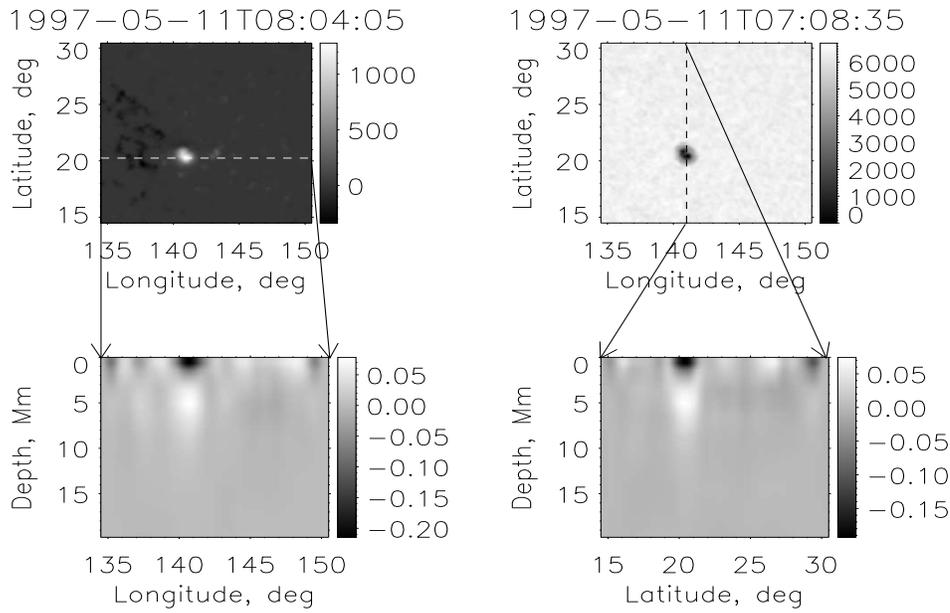}
\caption{Latitudinal and longitudinal cuts displayed on MDI magnetogram ({\em top left}) and continuum ({\em top right}) data
for NOAA 8038}
\end{center}
\end{figure*}
\begin{figure*}
\begin{center}
\includegraphics[height=12.5cm, width=8.cm, angle=90]{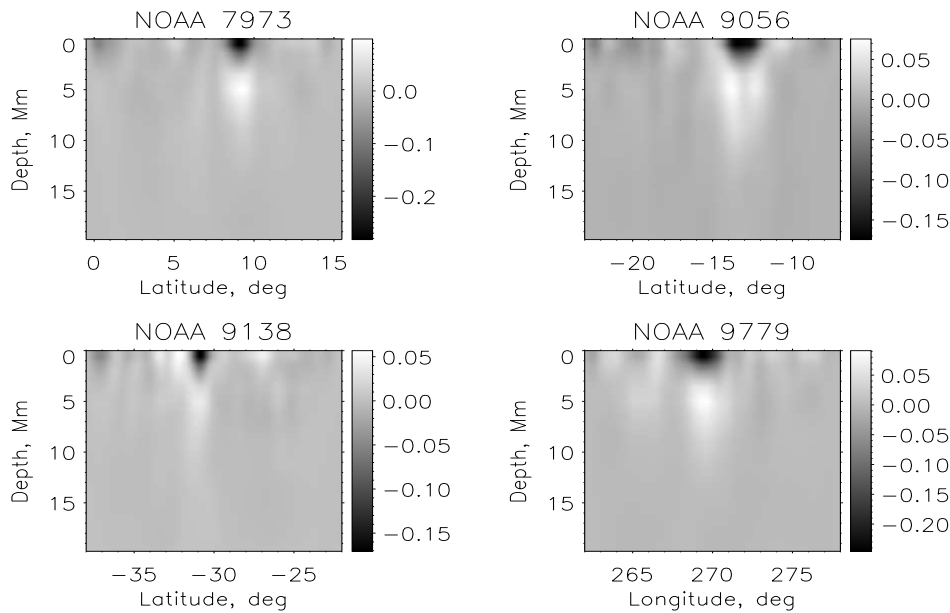} 
	\caption{Sound speed perturbation cut through the center of the sunspot 
	({\em as lower plots in Fig. 1}) for ({\em top, left to right}) NOAA 7973, NOAA 9056, ({\em bottom, left to right}) NOAA 9138 and NOAA 9779. }
\end{center}
\end{figure*}
The code for travel time inversion is based on a multi-channel deconvolution algorithm (MCD) \citepp{jensen98} enhanced by the addition of horizontal regularization. Since the sensitivity kernels used here are translationally invariant, the inversion can be considered as a convolution product. Thus, to speed up the inversion procedure, computations are performed in the Fourier domain. We discretise equation (\ref{eq_TT_Kernel}) following \citepp{jensen01}:
\begin{eqnarray}
d_i=\delta \tau ({\it \vec{k}}, \Delta_i ), \\
G_{ij}=K({\it \vec{k}}, z_j; \Delta_i ), \\
m_j=\delta s({\it \vec{k}}, z_j).
\end{eqnarray}
For each horizontal wavevector ${\it \vec{k}}$ we find the vector ${\it \vec{m}}$ that solves
\begin{eqnarray}
{\rm min} \{ \Vert (\vec{d}-G \vec{m})  \Vert^2_2 +\epsilon(\vec{k})^2 \Vert L\vec{m} \Vert^2_2 \},
\end{eqnarray}
where $L$ is a regularisation operator. In this work we take $L={\rm diag}(s_{0, 1}^{-1}, \ldots, s_{0, n}^{-1}),$ where $s_{0, i}$ the reciprocal of the sound-speed in the $i$-th layer of the reference model. 
We take $\epsilon(\vec{k})^2=\epsilon_0^2 (1+|\vec{k}|^2)^p$ with $p=70$ and $\epsilon_0^2=10^4$.
We invert for 14 layers in depth located at (.36, 1.17, 2.11,
					3.28, 4.74, 6.41,
					8.6 11.22, 14.28,
					17.78, 21.79, 26.31,
					31.41, 36.95) Mm.

\section{Discussion}
The results of sound-speed inversions are presented in Figures 1-5. We define the point of maximum absolute flux in the MDI magnetogram as the center of the sunspot for each of the features considered here. The top two panels in
Figure 1 show the remapped SOHO/MDI magnetogram and continuum images for NOAA 8038 with marked lines passing through the center of the sunspot representing longitudinal and latitudinal 
sound-speed perturbation depth cuts. The lower two panels present the actual depth cuts through these lines. Figure 2 contains selected sound-speed perturbation maps for some of the other Active Regions considered here. Figures 3 and 4 show sound-speed variation at two selected depths for each of the regions, with contours of the magnetic lines at 30 and 70 percent of the maximum absolute magnetic flux overplotted.
Figure 5 shows plots of sound-speed variation with depth at the center of each sunspot as defined above. 

Table 2 contains  extracted results of the inversion and feature-tracking for each of the regions under consideration. The first column contains NOAA number and
Carrington Rotation number, the second column presents the history of the region, e.g. NOAA 9056 was observed as NOAA 9017 one rotation earlier, while 
nothing was observed at the location of NOAA 7973 previously. The last column in the table contains additional information regarding further development of the Active Region. For example, NOAA 8038 soon after the time of observation develops into a bipolar $\beta \backslash \beta$ Active region, while NOAA 9056 was observed as NOAA 9096 one rotation later. We note from Tables 1 and 2 that the regions we describe as new are observed during the minimum and decaying phases of the solar cycle.

The results presented in the figures are in line with the earlier works by Kosovichev {\em et al} (2000), Couvidat {\em et al} (2005) and others. That is,  
we see a region of reduced sound speed directly underneath the sunspot, with a region of positive sound-speed perturbation situated at greater depth.
From Figures 3 and 4 we can see that these regions appear to be situated directly underneath each sunspot for the selected depths of $0.36$ and $6.41$ Mm. 

Analysis of the data in Table 2 suggests that sunspots associated with the new Active Regions, i.e. the ones that appeared within one Carrington Rotation, show larger amplitude of sound-speed perturbation, $\delta c / c$, than the Active Regions that were at least as old as one Carrington rotation period. The difference is more pronounced for the regions of reduced sound-speed profile.
Also, the extent of the region of reduced sound speed situated directly underneath the sunspot at the point of maximum magnetic flux tends to be larger for new sunspots. The validity of such sound speed anisotropy below the solar surface may have to be further verified by applying an umbra mask test (also known as cookie cutter test) which excludes observations taken in strongly magnetized regions (\cite{Zhao}, \cite{Hughes} and \cite{Korz}). 
Also, we note that the forward problem is currently modelled based on the basic assumption that magnetic field only modifies the effective sound speed, while the true picture is likely to be more complicated and would include showerglass effect, suppression and scattering of p-modes and other effects. 

\begin{table*}
\begin{center}
	\begin{tabular}{|c|c|c|c|c|c|c|}
	\hline
	NOAA \# &  CR \# & AR History & 0-crossing & min($\delta c /c$) &  max($\delta c /c$) & Comment \\
	\hline
	7973 & 1910   & new & 2.47 &  -0.282 &  0.106 &  \\
	8038 & 1922 & new &  2.11 &  -0.219 &  0.091 &appearing \\
	9056 & 1964 &  NOAA 9017 CR1963 & 2.12 & -0.176 & 0.078 & \# 9096CR1965 \\
	9138 & 1966 &  NOAA 9100 CR1965& 2.06 &  -0.177 &  0.073 & \\
	9142 & 1966 & NOAA 9105 CR1965 & 2.03 & -0.141 & 0.077 &\# 9201CR1968\\
	9779 & 1985 & new &  2.35 & -0.252 & 0.094 &  \\
	\hline
\end{tabular}
\caption{Properties of the Active Regions. The zero-crossing depth, i.e. the depth where $\delta c / c$ changes from negative to positive, 
is given in megameters.}
\end{center}
\end{table*}

\section{Conclusions}
In the present work, using the methods of time-distance helioseismology, we have looked at six different isolated sunspots at different stages of solar cycle
evolution. Using European Grid of Solar Observations' Solar Feature Catalogues and NOAA data we have extracted a number of parameters and tracked the history of each of the Active Regions.
We have extracted the sound speed perturbation at the time of observation as a function of depth directly underneath each region using Rytov-approximation sensitivity kernels.  We have used the inversion based on a modified Multi-Channel Deconvolution method with regularization chosen to be the inverse of the mean sound speed at each depth layer.

Our preliminary results indicate systematic differences in subsurface structure of the sunspot depending on the stage of its evolution or possibly the phase
of the solar cycle. Further investigations have to be carried out in order to verify these results. These include
making use of sensitivity kernels computed to take into account the specific details of our travel-time
computations and filtering procedure.
Inversion of travel-time data is an ill-posed problem and further improvements to the procedure can be made by including a noise covariance matrix and experimenting with different choices of regularization operator. 

In addition, further work is to be carried out to verify these results by investigating a statistically significant number of sunspots.
\acknowledgements
We thank J. Jensen for helping with code development and kindly providing the sound-speed sensitivity kernels used in this work.
\begin{figure*}
\begin{center}
\begin{tabular}{ccc}
\hfill
\includegraphics[height=3.2cm]{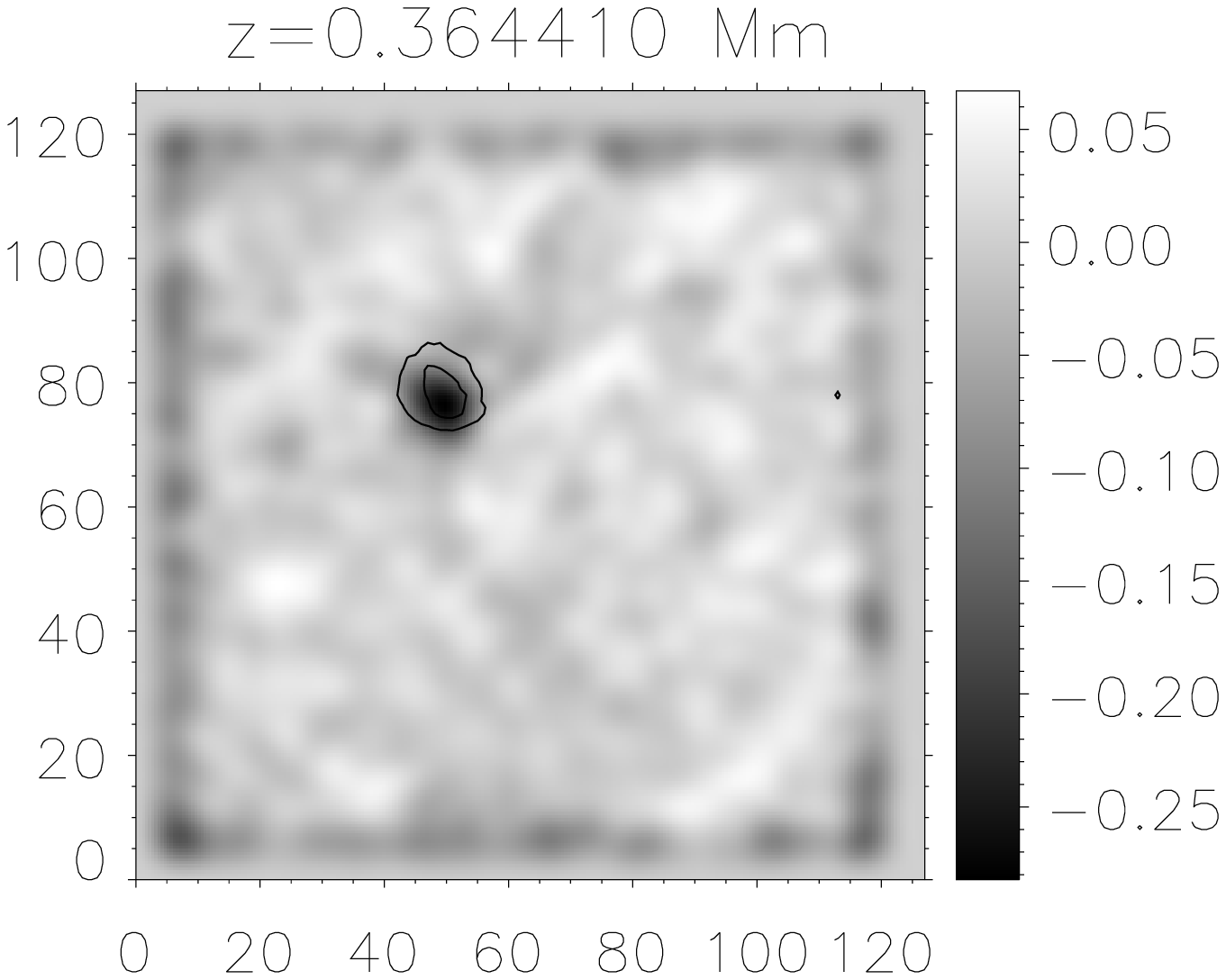} & 
\includegraphics[height=3.2cm]{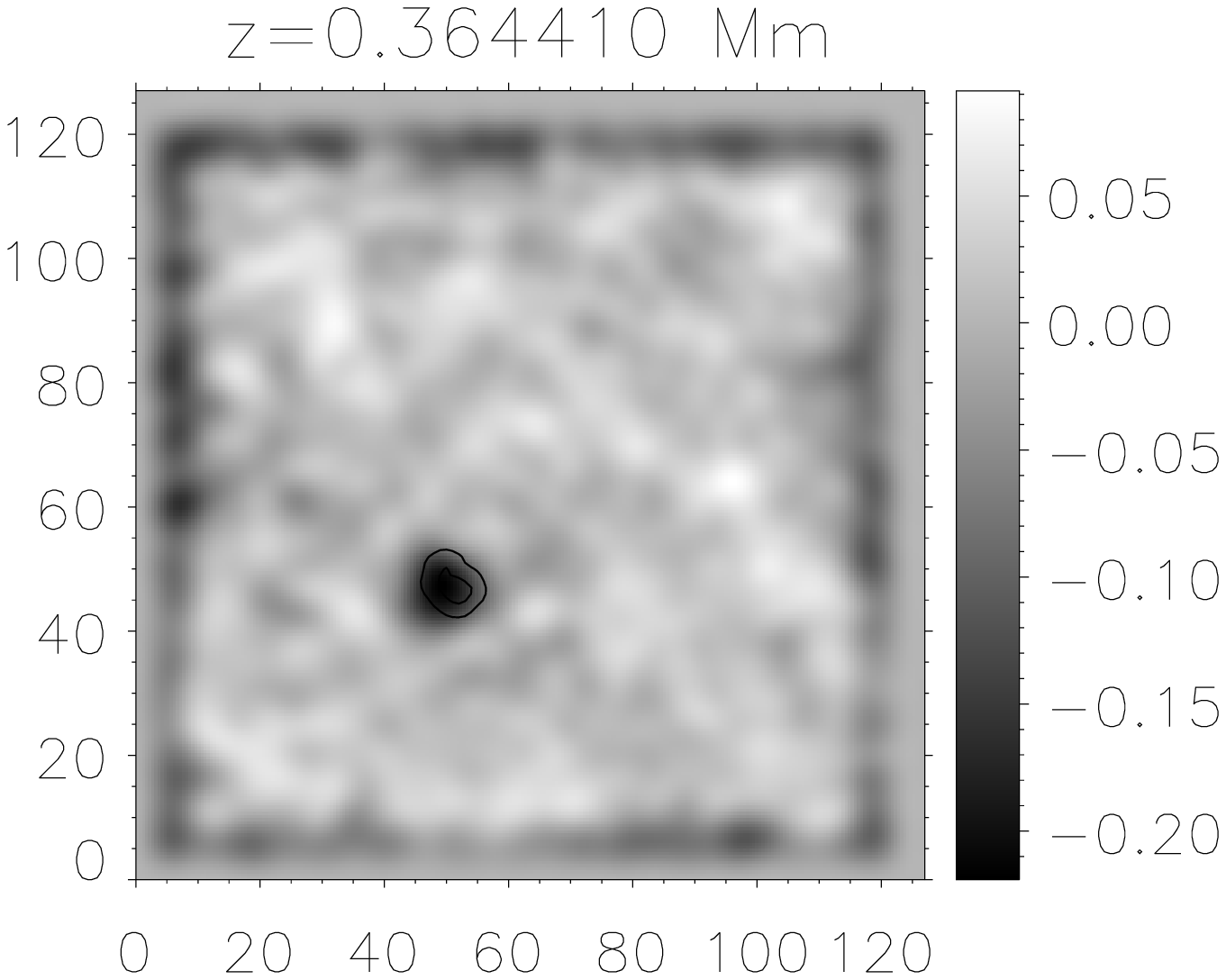} & 
\includegraphics[height=3.2cm]{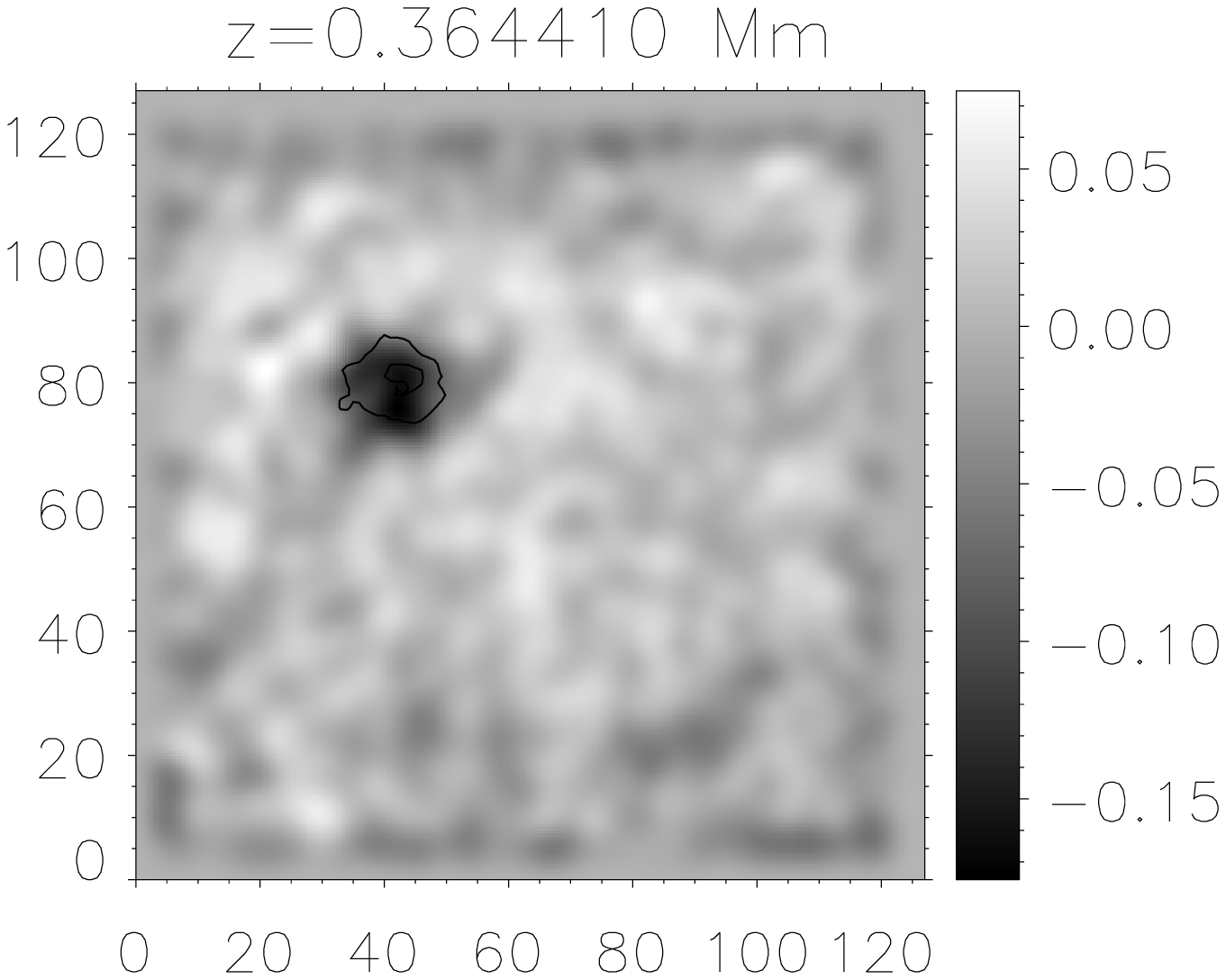} \\
\includegraphics[height=3.2cm]{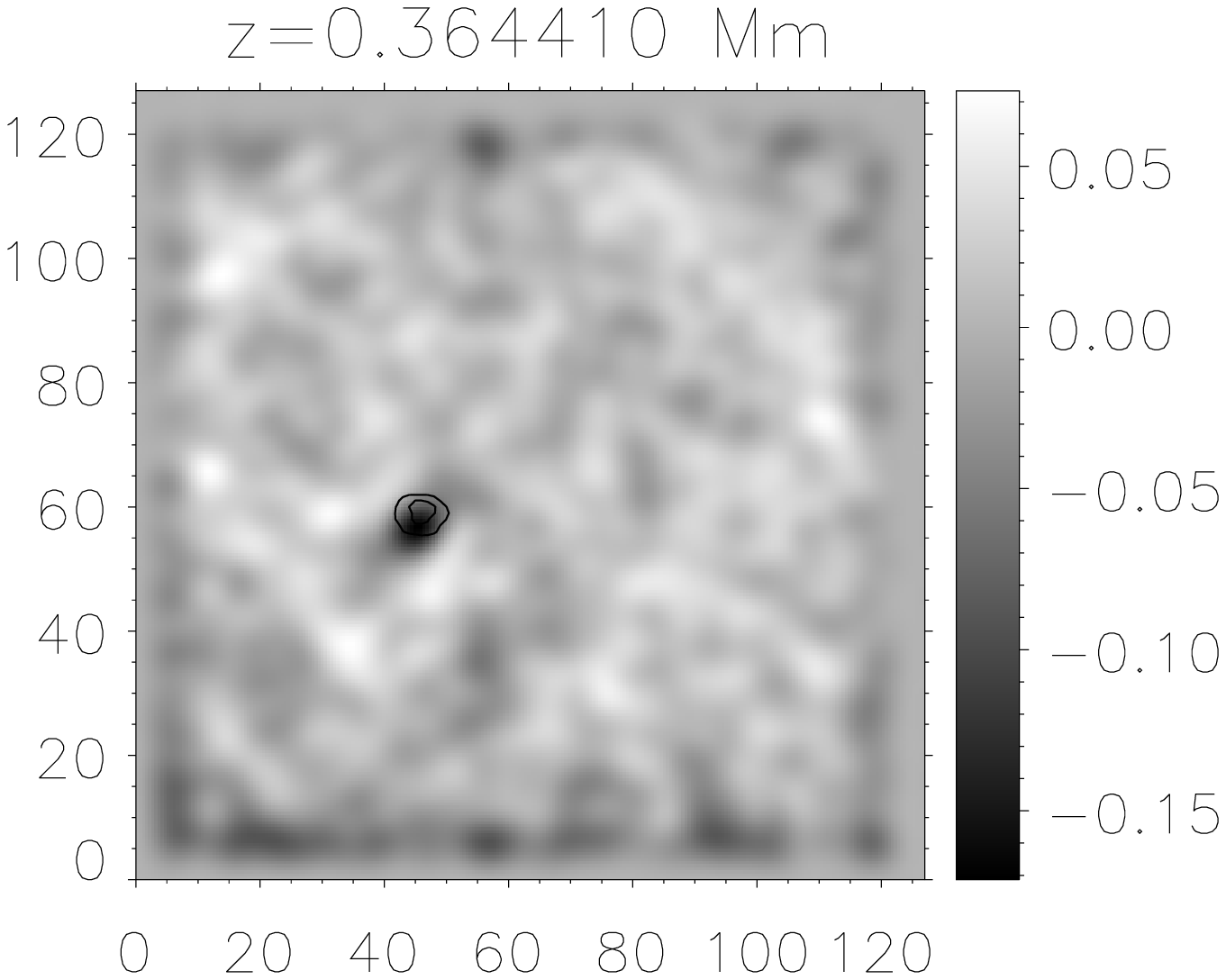} &
\includegraphics[height=3.2cm]{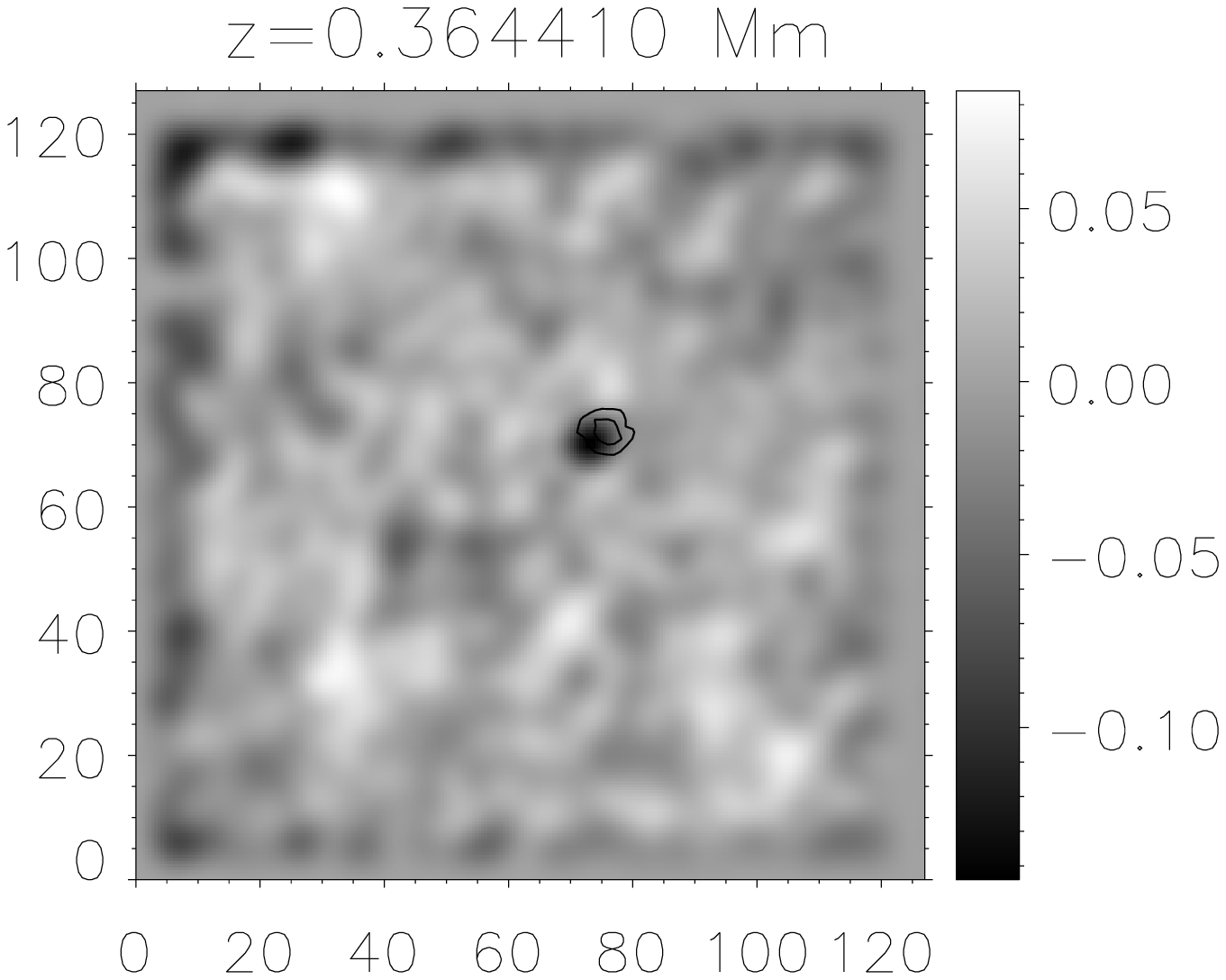} &
\includegraphics[height=3.2cm]{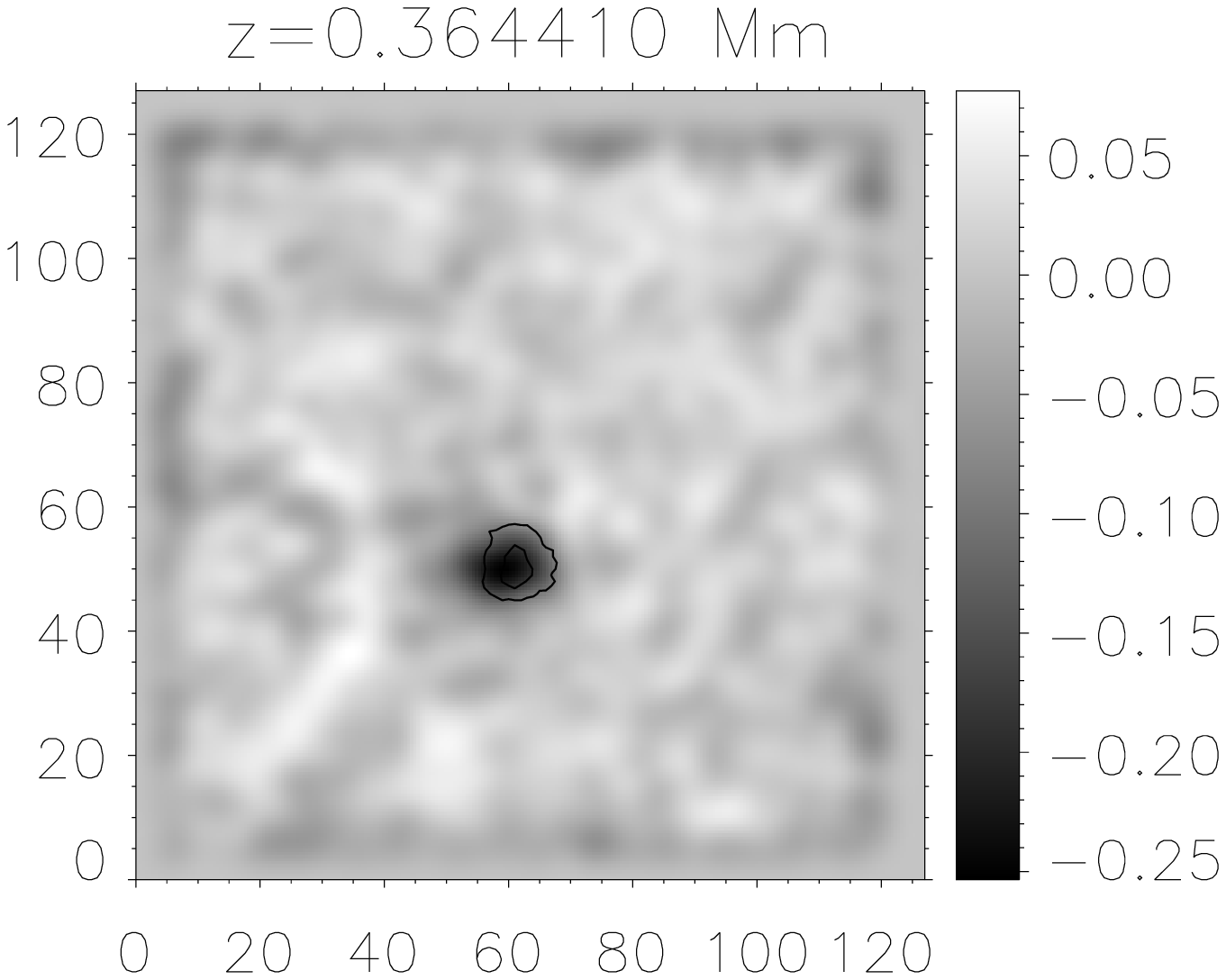}
 \end{tabular}
 \caption{Sound-speed perturbation cut at $z=0.36 {\rm Mm}$ for ({\em top row, left to right}) NOAA 7973, NOAA 8038, NOAA 9056, ({\em bottom, left to right}) NOAA 9138, NOAA 9142, NOAA 9779.}
 \end{center}
\end{figure*}

\begin{figure*}
\begin{center}
\begin{tabular}{ccc}
\hfill
\includegraphics[height=3.2cm]{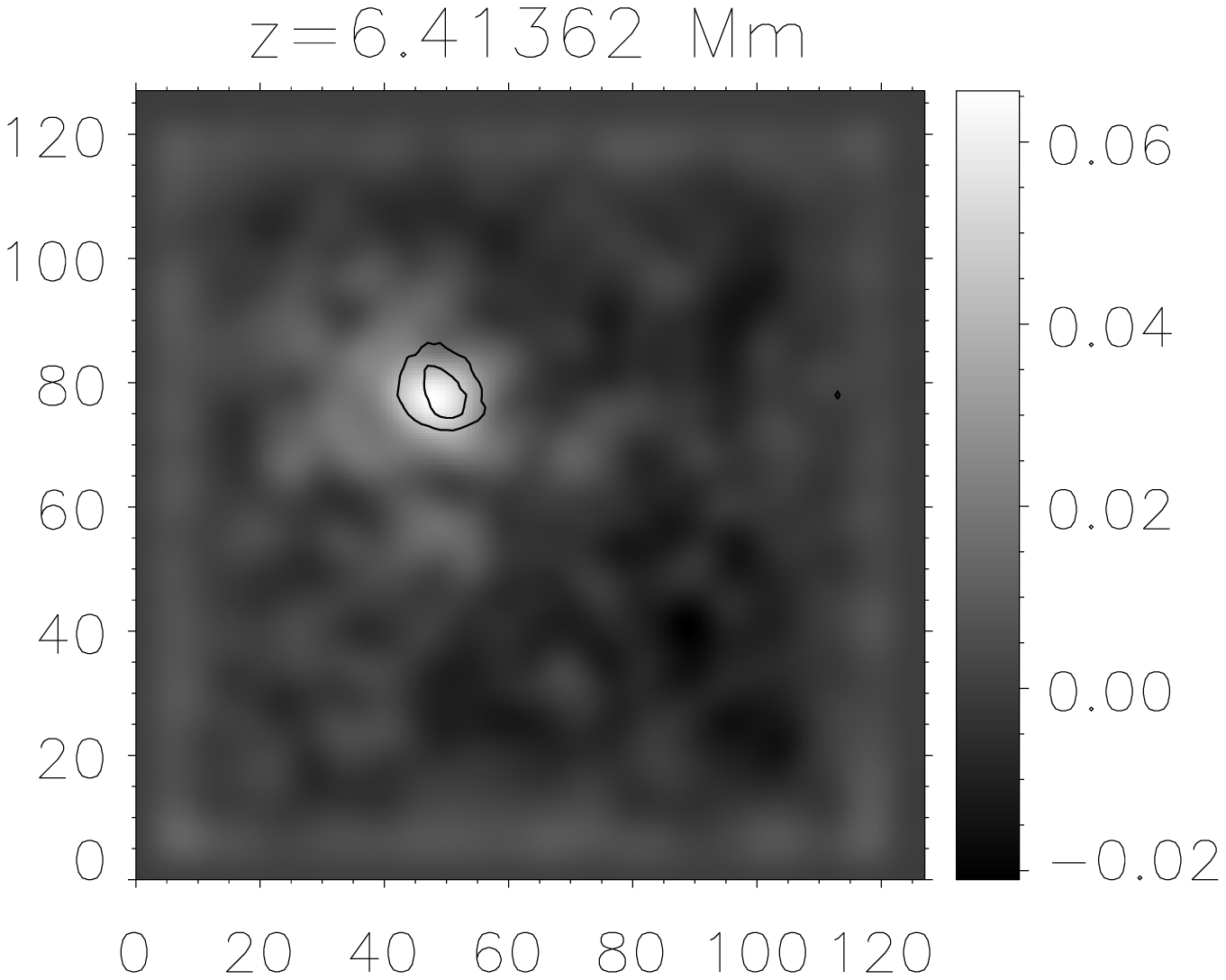} & 
\includegraphics[height=3.2cm]{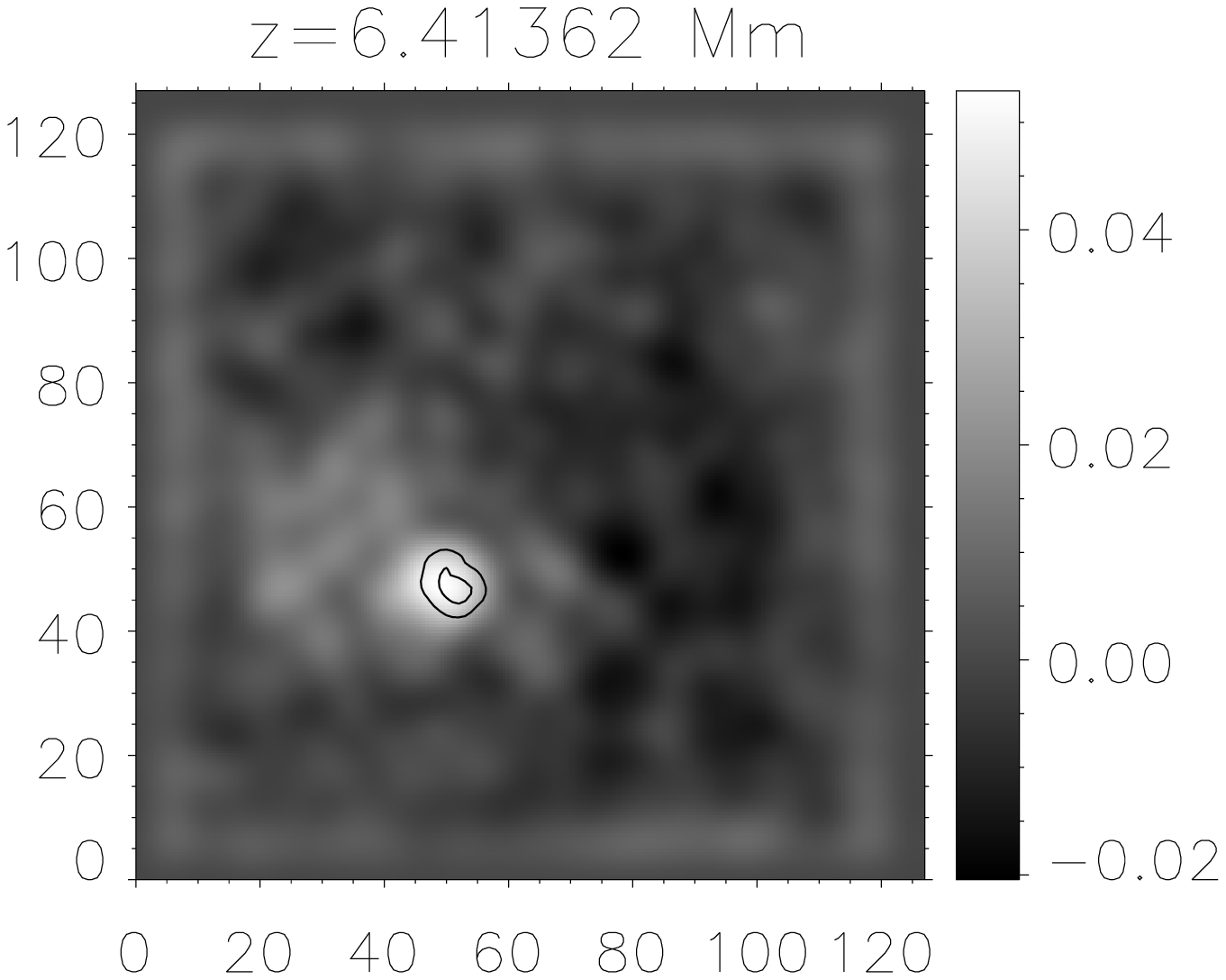} &
\includegraphics[height=3.2cm]{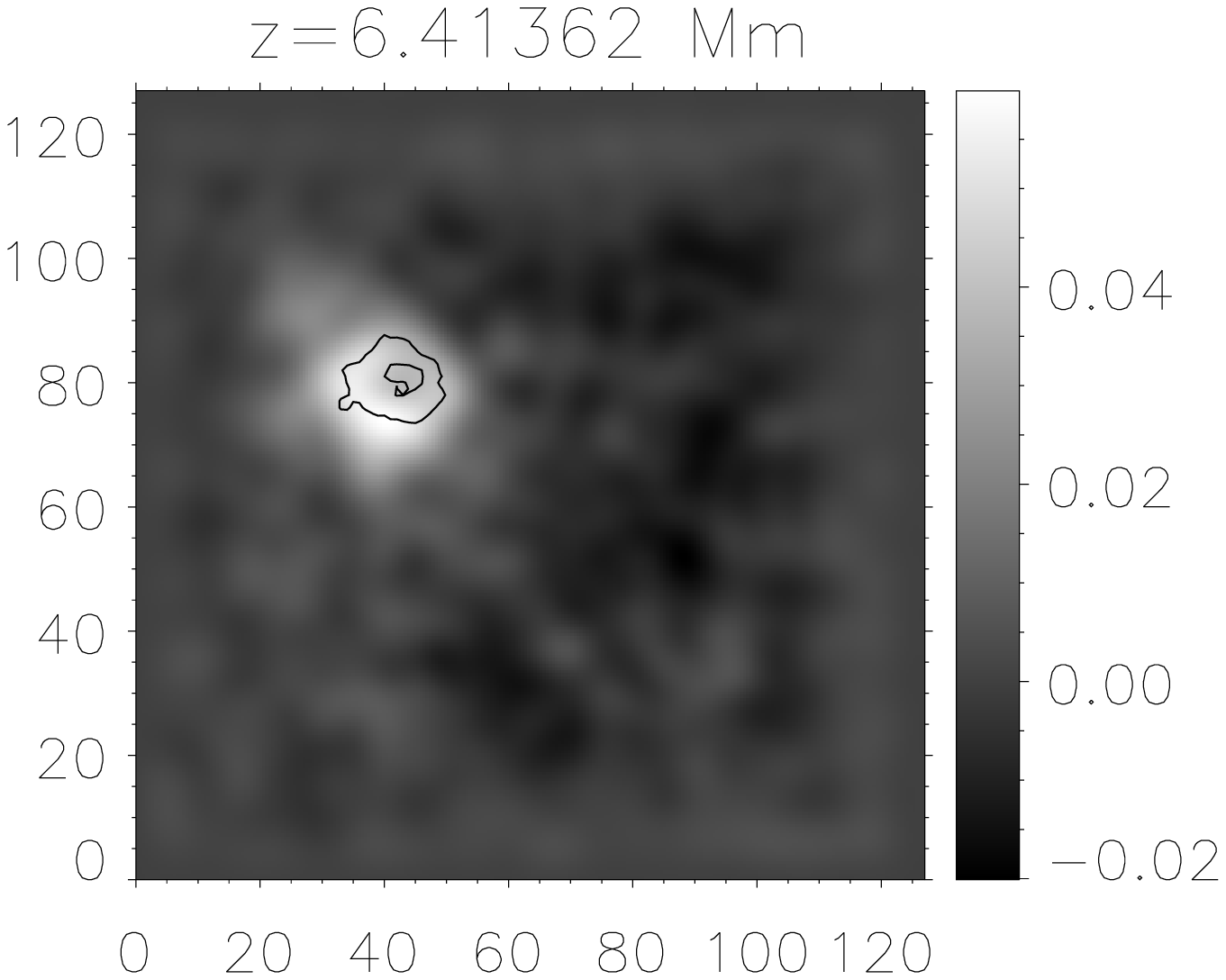} \\
\includegraphics[height=3.2cm]{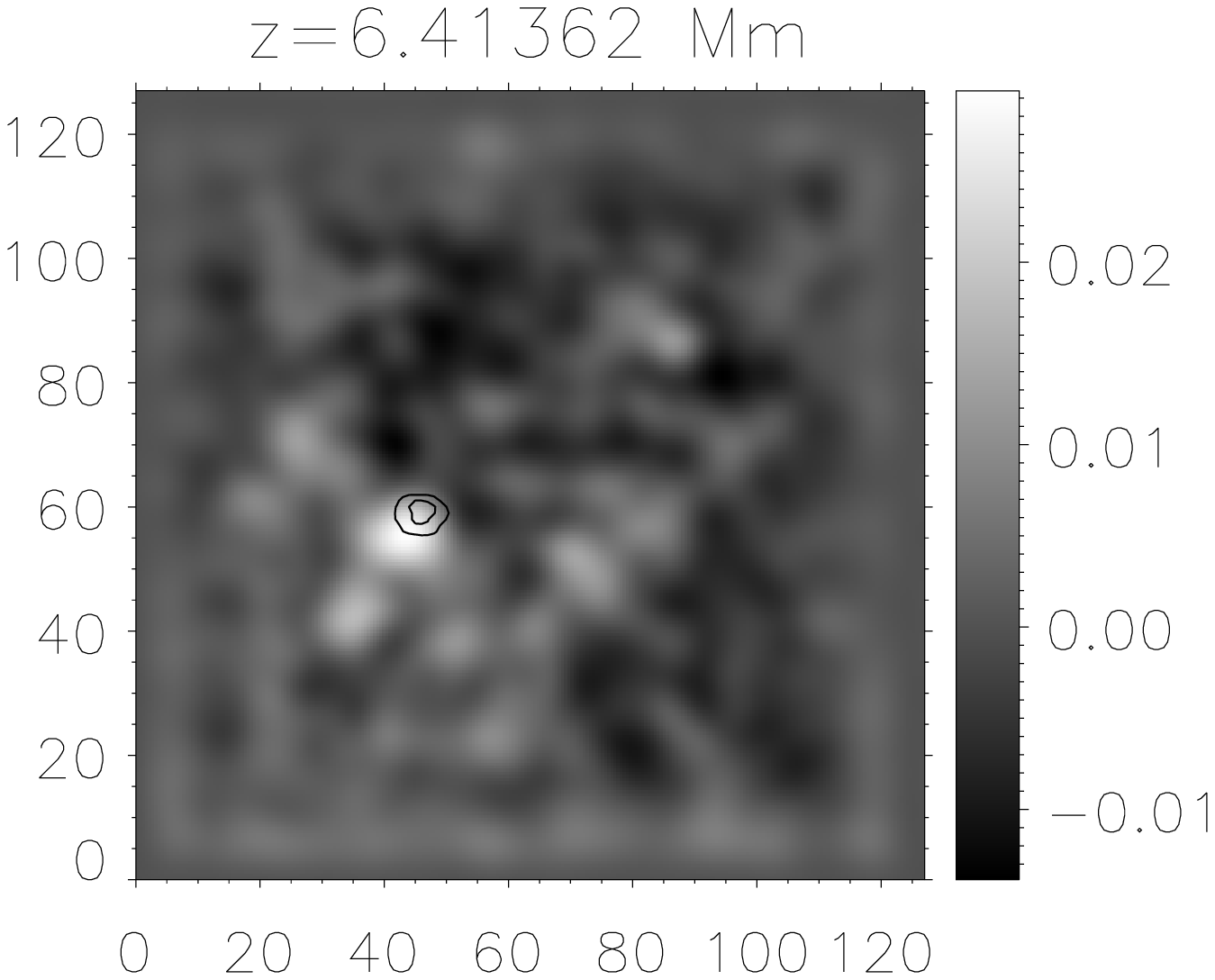} &
\includegraphics[height=3.2cm]{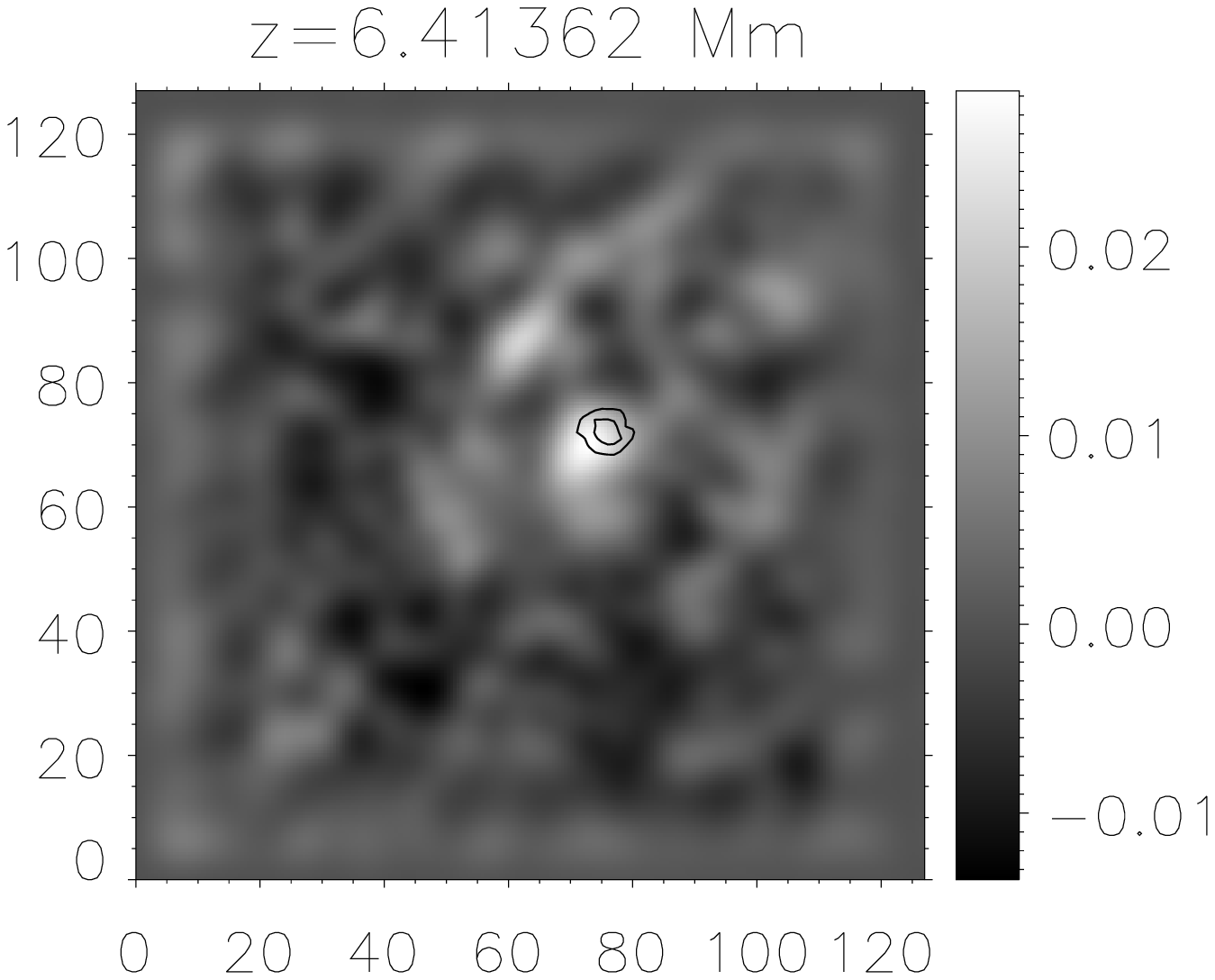} &
\includegraphics[height=3.2cm]{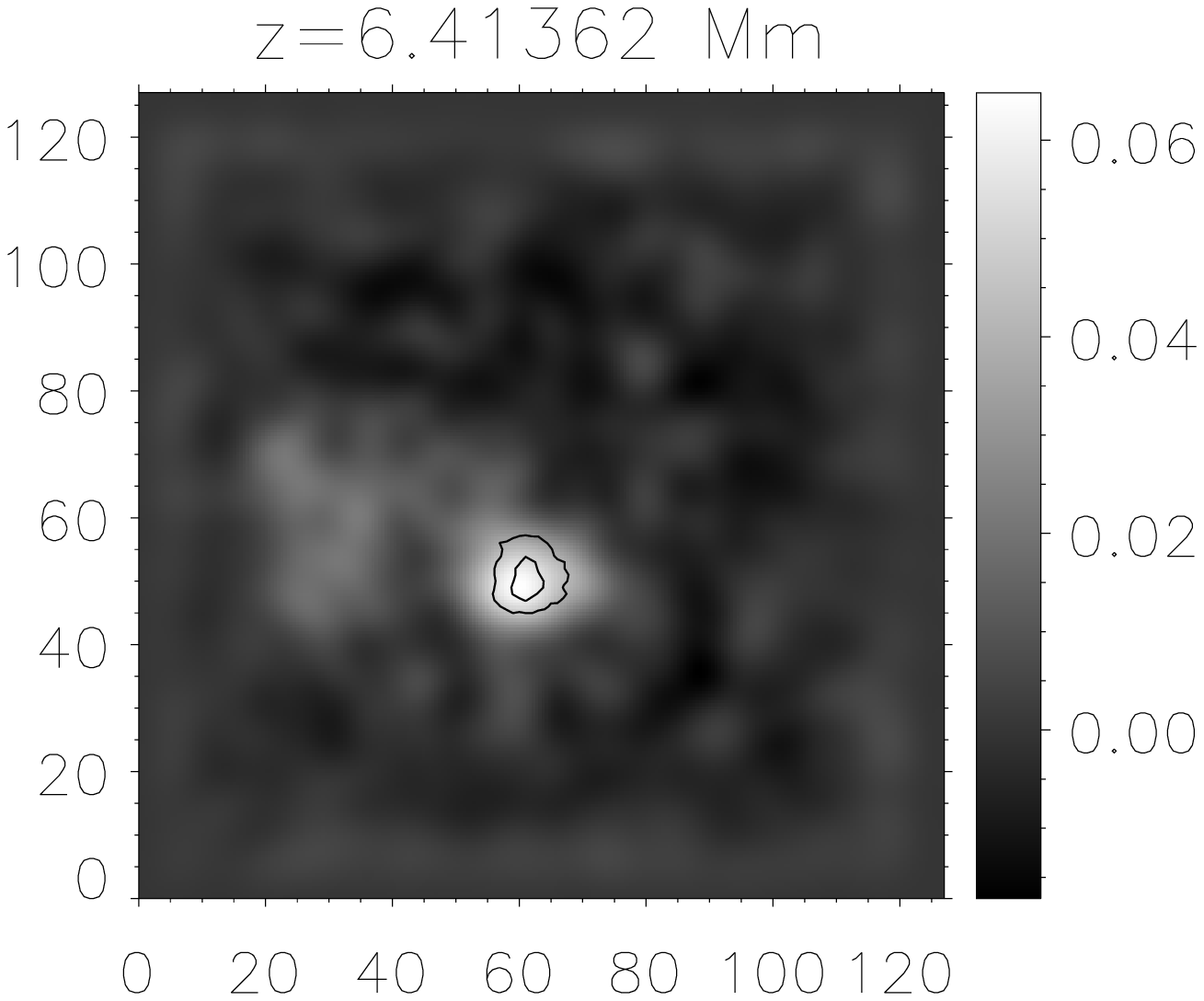} 
\end{tabular}
\caption{Sound-speed perturbation cut at $z=6.41 {\rm Mm}$ for ({\em top row, left to right}) NOAA 7973, NOAA 8038, NOAA 9056 and ({\em bottom, left to right}) NOAA 9138, NOAA 9142, NOAA 9779.}
\end{center}
 \end{figure*}
 
\begin{figure*}
\begin{center}
\includegraphics[width=6.5cm, height=14.5cm, angle=90]{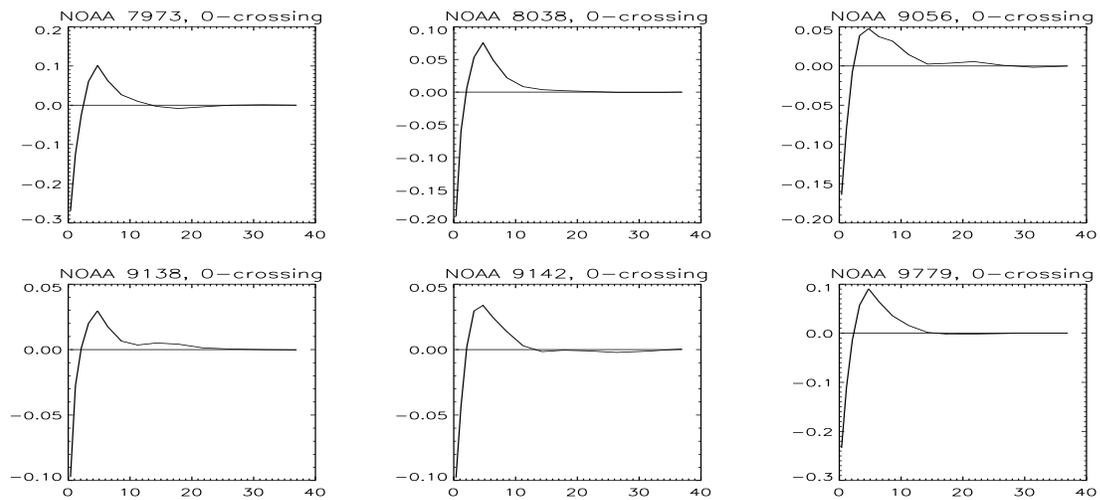}
\caption{Sound-speed perturbation, as a function of depth, at the center of the sunspot defined as the point of maximum absolute magnetic flux.}
\end{center}
\end{figure*}

\bibliographystyle{plain}

\end{document}